\documentclass[5p,sort&compress]{elsarticle}
\usepackage{amsmath}            
\usepackage{wasysym }           
\usepackage{graphics}          
\newcommand{\sfrac}[2]{\frac{#1}{#2}}
\newcommand{\V}[1]{\mathbf{#1}}
\renewcommand{\deg}{$^\circ$}
\newcommand{\mc}{\multicolumn}
\newcommand{\B}{\overline}

\begin{document}

\begin{frontmatter}
\title{Gains from the upgrade of the cold neutron triple-axis spectrometer FLEXX at the BER-II reactor}

\author[label1]{M. D. Le}
\author[label1]{D. L. Quintero-Castro}
\author[label1]{R. Toft-Petersen}
\author[label1]{F. Groitl}
\author[label1,label2]{M. Skoulatos}
\author[label1,label3]{K. C. Rule}
\author[label1]{K. Habicht}
\address[label1]{Helmholtz-Zentrum Berlin f\"ur Materialen und Energie, Hahn-Meitner-Platz 1, D-14109 Berlin, Germany}
\address[label2]{Laboratory for Neutron Scattering, Paul Scherrer Institut, CH-5232 Villigen PSI, Switzerland}
\address[label3]{The Bragg Institute, ANSTO, Kirrawee DC NSW 2234, Australia}

\date{\today}


\begin{abstract}

The upgrade of the cold neutron triple-axis spectrometer FLEXX is described. We discuss the characterisation of the gains from the new primary
spectrometer, including a larger guide and double focussing monochromator, and present measurements of the energy and momentum resolution and of the
neutron flux of the instrument. We found an order of magnitude gain in intensity (at the cost of coarser momentum resolution), and that the incoherent
elastic energy widths are measurably narrower than before the upgrade. The much improved count rate should allow the use of smaller single crystals
samples and thus enable the upgraded FLEXX spectrometer to continue making leading edge measurements.

\end{abstract}


\end{frontmatter}


\section{Introduction} \label{sec-intro}

The cold neutron triple axis spectrometer FLEX~\cite{vorderwisch_flex_95} has had over 15 years of successful operation, serving both as a platform
for the development of the neutron resonance spin echo (NRSE) technique~\cite{habicht_nrse_04} and as a work-horse instrument for magnetic neutron
scattering~\cite{dmitri_tbfeo3,toftpetersen_linipo4,diana_sr2cr2o8,coldea_ising_e8,thielemann_spinladder}. Recently, the primary spectrometer was
completely rebuilt with new $m=3$ guides including a converging elliptical section to focus neutrons onto a virtual source. The neutrons are
subsequently imaged onto a new double focussing monochromator, ensuring an increase in neutrons reaching the sample~\cite{skoulatos_flexx}. In
addition, a new velocity selector is used to remove higher order scattering which eliminates the need for filters. A polarising S-bender may be
translated into the beam before the elliptical guide section where the beam is relatively well collimated, allowing the gains from focussing neutrons
onto the sample to be realised for polarised measurements also. Furthermore new compact NRSE arms with new coils and shielding~\cite{felix_thesis}
have been constructed. The improved coil design permits a larger beam cross-section to be transmitted, and also larger coil tilt angles to be reached
allowing measurements of steeper dispersions. 

In this article we present a characterisation of the instrument, including the gains in count rate and measured resolution.  The primary spectrometer
upgrade is described in section~\ref{sec-primary}, including the performance of the velocity selector and monochromator from measurements at the
sample position and comparison with Monte Carlo simulations. In section~\ref{sec-gains} the energy and momentum resolution of the instrument and the
gains in count rate compared with that before the upgrade are investigated using measurements of single crystals of lead and silicon.

\section[The Primary Spectrometer]{The Primary Spectrometer\protect\footnote{A\MakeLowercase{ll measurements in this section were taken with a 3~cm virtual source.}}} \label{sec-primary}

\begin{figure*}
  \begin{center}
    \includegraphics[width=0.99\textwidth,bb=36 162 810 479]{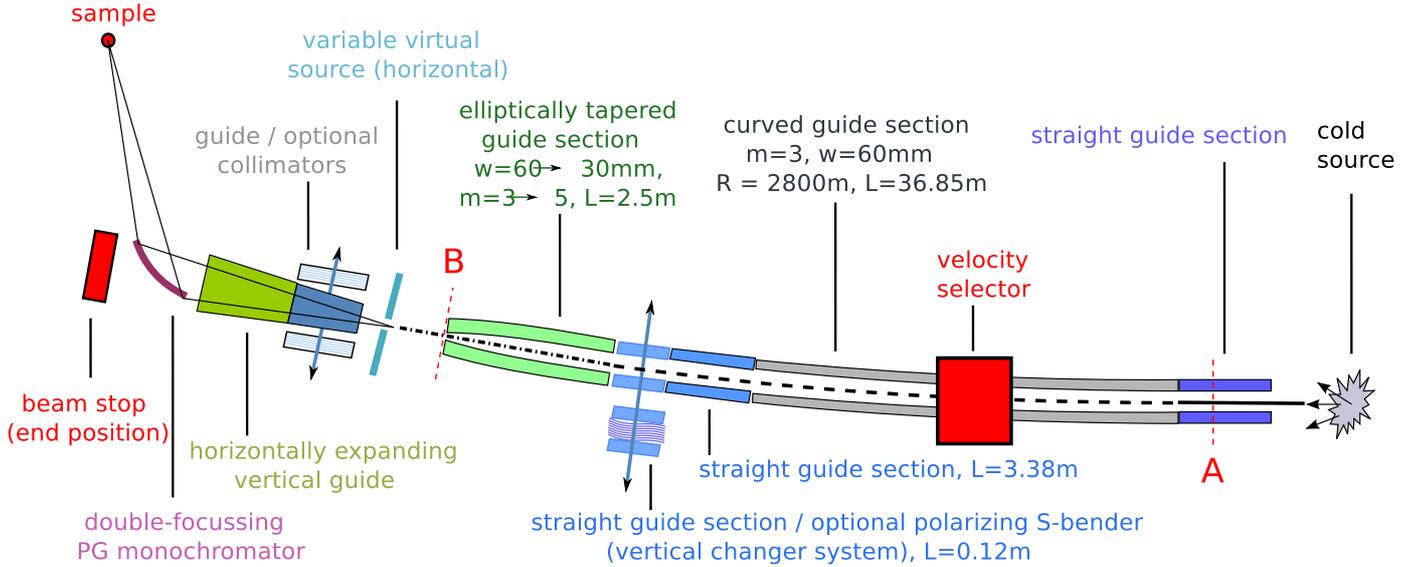}
    \caption{Schematic drawing of the NL1B guide. The dashed red lines labelled "A" and "B" are positions for the gold foil measurements discussed in the text.}
    \label{fg:guideschematic} \end{center}
\end{figure*}

A schematic drawing of the new 50~m long NL1b guide, which is dedicated to the FLEXX instrument, is shown in figure~\ref{fg:guideschematic}. Beginning 
at the upgraded cold source, there is a common extraction section followed by a short (1.5~m) straight guide. At the end of this, 5~m away from the source,
an 18~m curved guide leads to a neutron velocity selector. The curved guide continues after the selector for another 18~m leading to a short (3.5~m) 
straight guide section, ensuring a more uniform beam~\cite{habicht_skoulatos_virtualsource}. A compact S-bender polariser mounted on a vertical translation stage is installed after this guide, and is followed by an
elliptical guide which focusses neutrons onto a horizontal virtual source. Immediately after the virtual source is a horizontal translation stage with
two collimators (40' and 60') and a short guide section ("open" collimation). Neutrons from the virtual source are allowed to spread in the subsequent
expanding section, in which only the top and bottom surfaces are neutron reflecting, before entering the monochromator section. Due to constraints
imposed by the guide shielding, the expanding section is not evacuated, but is rather filled with a slight overpressure of helium gas.

\begin{figure*}
  \begin{center}
    \includegraphics[width=0.49\textwidth,bb=19 180 600 609]{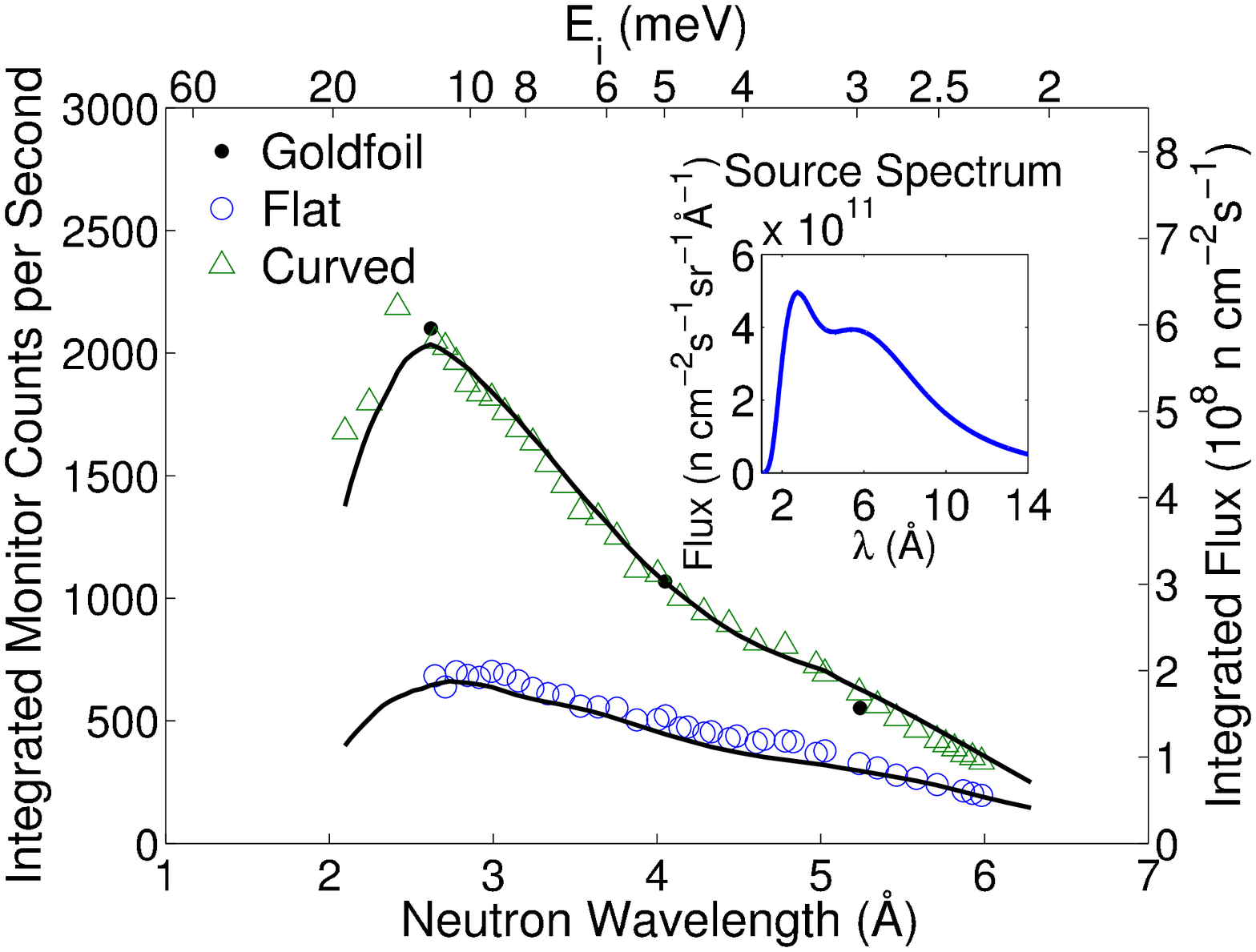}
    \includegraphics[width=0.49\textwidth,bb=19 180 600 609]{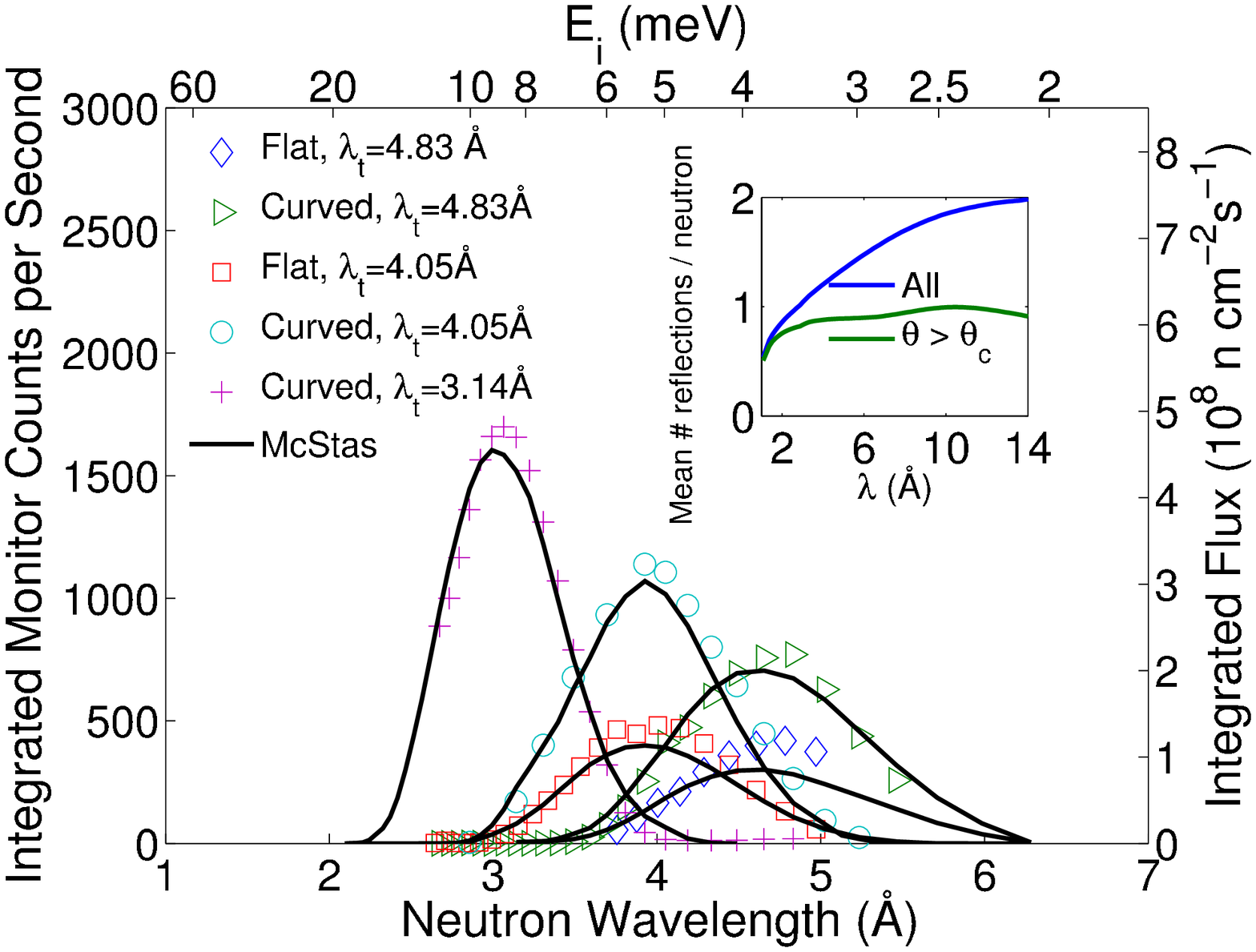}
    \caption{Fitted integrated intensities of monochromator rocking ($\theta_m$) scans as a function of incident wavevector.
             The left panel shows the measured data with the monochromator flat (blue circles) and optimally focussed (green triangles). The neutron
             flux at the sample position measured by neutron activation in gold foils is shown as solid black circles with the right hand scale.
             McStas simulations using the spectrum shown in the inset and described in the text are shown as solid lines.
             The right panel shows the measurements with the velocity selector fixed at particular speeds to allow transmission of neutrons with
             incident wavevector around 1.3~\AA$^{-1}$ (transmitted wavelength around $\lambda$=4.83~\AA, blue diamonds for flat and green triangles 
             for curved monochromator), 1.55~\AA$^{-1}$ ($\lambda$=4.05~\AA, red squares for flat monochromator, light blue circles for curved) and 
             2~\AA$^{-1}$ ($\lambda$=3.14~\AA, purple crosses, curved monochromator). The inset to this figure shows the average number of
             guide reflections as a function of neutron wavelength.}
    \label{fg:spectrum} \end{center}
\end{figure*}

In order to characterise the performance of the guide system and velocity selector, a series of monochromator rocking ($\theta_m$) scans with a low
efficiency neutron monitor at the sample position were made for a variety of monochromator scattering angles ($2\theta_m$) and velocity selector
speeds. The neutron monitor was masked by a borated polyethylene sheet such that only a 1.5~cm diameter circular window is visible so as to better represent
a similarly sized sample.
The area of the peaks measured with the monochromator optimally curved (green triangles) and flat (blue circles) is shown in the left panel of
figure~\ref{fg:spectrum}. The data was corrected for scattering due to 1~cm of aluminium in the windows between the guide sections, and for the
efficiency of the monitor, which was assumed to be inversely proportional to the neutron velocity, normalised to a velocity of 2.2~km~s$^{-1}$. The
solid black lines are the result of a McStas~\cite{mcstas} simulation of the instrument, using the source spectrum shown in the inset. This source is
modelled using two Maxwellian distributions, with temperatures $T_1=$60~K and $T_2=$11~K, and flux $I_1=$1.24$\times
10^{12}$~ncm$^{-2}$s$^{-1}$sr$^{-1}$\AA$^{-1}$ and  $I_2=$0.61$\times 10^{12}$~n cm$^{-2}$s$^{-1}$sr$^{-1}$\AA$^{-1}$.

Gold foil absorption measurements at the sample position for three incident wavevectors, $k_i$=1.2, 1.55 and 2.4~\AA$^{-1}$, with the monochromator
focussed are shown as solid black circles in the figure, and are in good agreement with the rocking scans. The flux derived from these measurements,
shown by the scale on the right was used to determine the flux parameters of the McStas source. These were further checked against the results of gold
foil absorption measurements of the white beam flux which are undertaken every six months as part of a program to measure the quality of the newly
installed guides. The measurements are carried out at two positions indicated by dashed red lines in figure~\ref{fg:guideschematic} and are summarised
in table~\ref{tab:goldfoil}. The 10\% reduction in the measured flux was also correlated with reductions in the measured count rate from standard
calibration measurements with the neutron monitor. However, we note that this may be due to changes in the arrangement of the fuel elements in the
core, and not necessarily due to a degradation of the guide.

\begin{table} \renewcommand{\arraystretch}{1.3}
\begin{center}
  \begin{tabular}{@{\extracolsep{\fill}}r|c|c|c}
   \hline 
                        & April 2012         & October 2012       &   McStas              \\
   Rotary shutter & 7.87$\times$10$^9$ & 7.04$\times$10$^9$ &   10.67$\times$10$^9$ \\
   Monochromator & 2.48$\times$10$^8$ & 2.25$\times$10$^8$ &\ \ 2.36$\times$10$^8$ \\
   \hline 
  \end{tabular}
  \caption{White beam integrated neutron flux in n~cm$^{-2}$s$^{-1}$ derived from gold foil absorption measurements compared to McStas calculations.
           The position of the gold foil installed after the rotary shutter is labelled "A" in figure~\ref{fg:guideschematic} whilst the other is "B".
	   The much lower integrated flux at the end of the guide is due primarily to the velocity selector which only allows a relatively narrow
	   wavelength band to pass. The measurements and calculations were performed with the selector set to transmit neutrons with wavelengths
           around 4~\AA.} 
  \label{tab:goldfoil}
\end{center}
\end{table}

Whilst the deduced source spectrum shown as an inset to the left panel of figure~\ref{fg:spectrum} suggests that there should be a peak in the neutron
spectrum around 6~\AA, this is not apparent in the measurements at the sample position. This is due to the larger number of guide reflections these
neutrons encounter, so that their transmission is correspondingly less. Whereas 2~\AA~neutrons are transmitted through the guide with, on average,
half a reflection, 6~\AA~neutrons encounter three times as many. Moreover, many of these reflections are at angles higher than the critical angle of
$^{58}$Ni, with an attendant lower reflectivity. The mean number of guide reflections is shown as an inset to the right panel of figure~\ref{fg:spectrum}.

The performance of the velocity selector is shown in the right panel of figure~\ref{fg:spectrum}. The points in the graph represent the area of a
monochromator rocking scan with the velocity selector fixed at three particular speeds, corresponding to optimal (peak) transmission of neutrons with
wavevectors $k_i$=1.3, 1.55 and 2.0~\AA$^{-1}$ ($\lambda$=4.83, 4.05 and 3.14~\AA~respectively). The data is well modelled by the McStas simulations
(solid black lines). The model indicates that the peak transmission through the velocity selector is approximately 70\% at low speeds, below
$\approx$12~krpm corresponding to wavelengths above $\approx$4~\AA, falling linearly to approximately 40\% at the maximum speed of $\approx$26~krpm
corresponding to $\approx$2~\AA~for zero tilt. The twist angle of the blades, 19.7\deg, can be effectively varied by tilting the velocity selector.
Thus the selector is mounted on a goniometer which may be rotated by $\pm5$\deg vertically in order to access higher or lower wavelength neutrons. 
The simulations also indicate that the wavelength bandwidth of the velocity selector is approximately 30\% of the central wavelength, that is $\Delta\lambda/\lambda\approx 0.3$. 
The second order transmission
for 4~\AA~neutrons was measured using the forbidden (001) reflection of Pb to be less than 10$^{-4}$. Finally we note that the spectrometer may be
operated in second order mode by setting the velocity selector speed to be twice the calculated optimum for some particular monochromator setting,
since the speed is approximately inversely proportional to the neutron wavelength.

\begin{figure}
  \begin{center}
    \begin{tabular}{@{\extracolsep{\fill}}cc}
      \includegraphics[width=\columnwidth,bb=19 179 585 584]{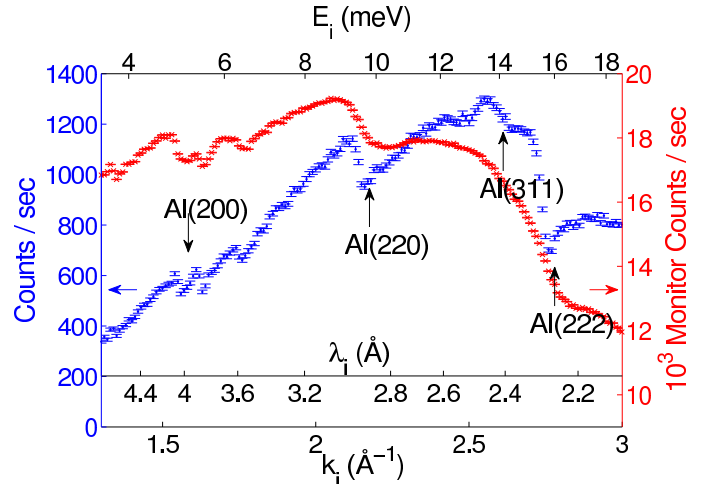}
    \end{tabular}
    \caption{Measured spectrum from a 1~cm diameter Vanadium rod, uncorrected for scattering from Aluminium windows in the guide. Blue dots (left
             scale) are detector counts per second, whilst red crosses (right scale) are from a low efficiency monitor placed after the monochromator.}
    \label{fg:vanspec} \end{center}
\end{figure}

Figure~\ref{fg:vanspec} shows the elastic incoherent scattering from a 1~cm vanadium rod as a function of the neutron energy, and gives an indication
of the energy ranges where the neutron flux at the sample position is maximised. Unlike the data in figure~\ref{fg:spectrum}, the count
rate here has not been corrected and clearly shows the effect of scattering from aluminium windows separating each guide sections. It is apparent that
the maximum intensity is around $E_i\approx$14~meV with a rather sharp drop at slightly higher energies due to an aluminium Bragg edge.

\begin{figure}
  \begin{center}
    \begin{tabular}{@{\extracolsep{\fill}}cc}
      \includegraphics[width=0.5\columnwidth,bb=62 199 540 595]{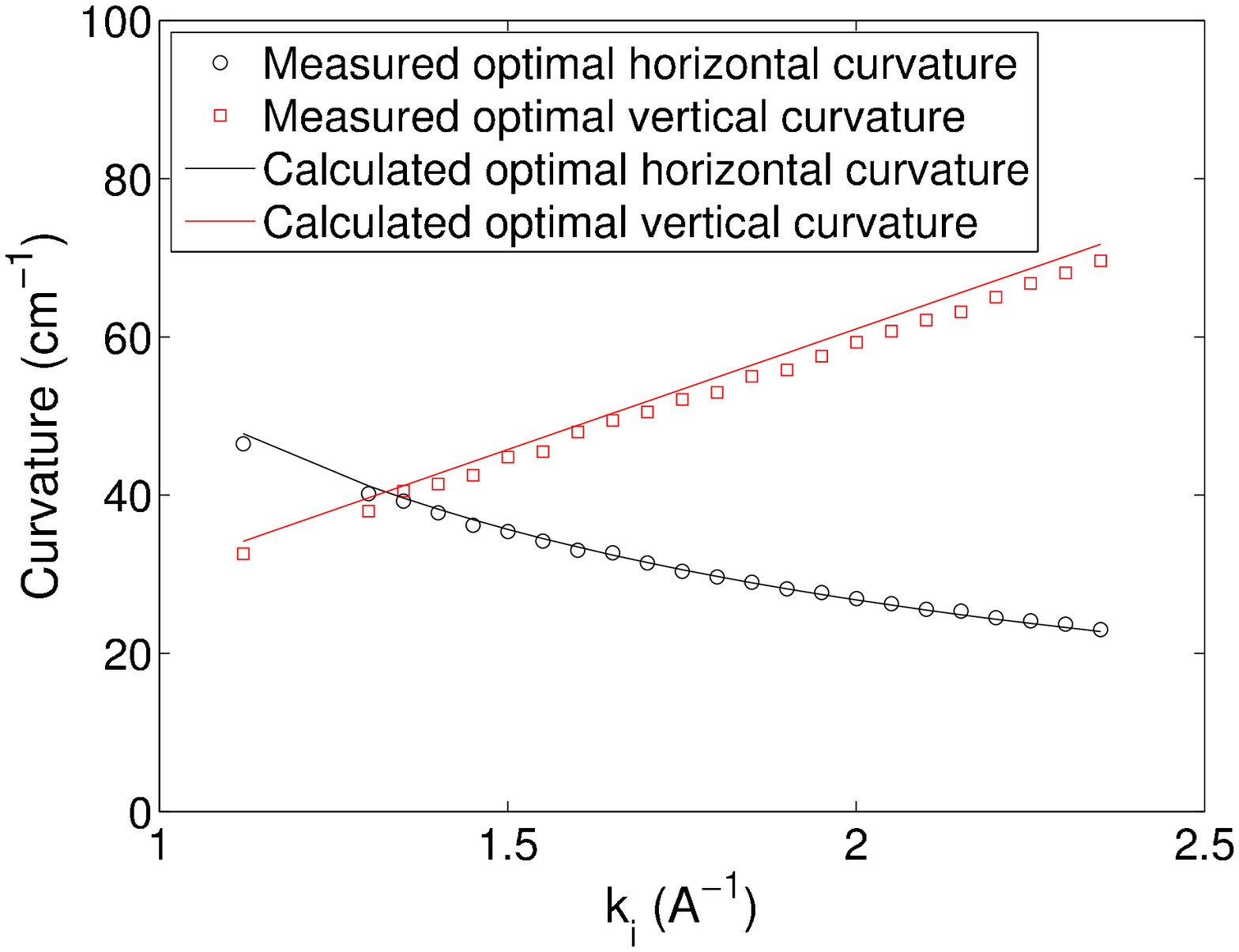} &
      \includegraphics[width=0.5\columnwidth,bb=62 199 546 584]{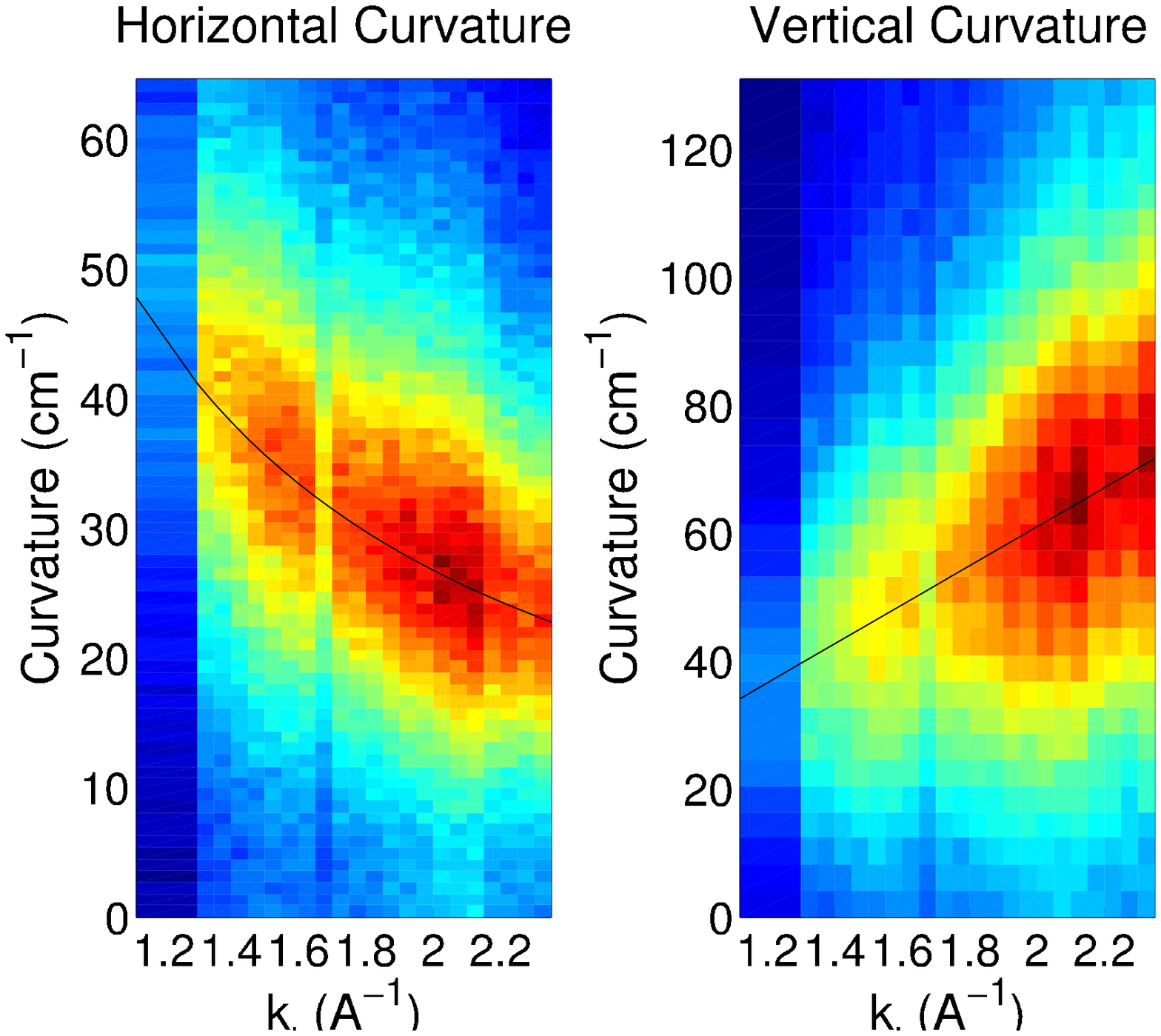}
    \end{tabular}
    \caption{Scans of the monochromator curvature onto a monitor mounted at the sample position. The left panel shows the
             fitted peak curvatures (red squares vertical, black circles horizontal) and the curvatures calculated from
	     equations~\ref{eq:rmh} and~\ref{eq:rmv} (lines) for distances $L_0=L_1=1.75$~m. The right panel show the scans
             as a colour contour plot with the calculated optimal curvature as solid black lines.}
    \label{fg:monocurvature} \end{center}
\end{figure}

We now turn to the characterisation of the double focussing monochromator. Scans of the horizontal and vertical curvature of the monochromator at
many different incident neutron wavevectors are summarised in figure~\ref{fg:monocurvature}. The loci of maximum intensities as a function of
monochromator take off angle, 2$\theta_m$, were found to agree well with analytical expressions for the optimum curvature derived from geometrical
optics,

\begin{equation} \label{eq:rmh}
\rho_h = \frac{(L_0+L_1)\sin\theta_m}{2 L_0 L_1}
\end{equation}

\begin{equation} \label{eq:rmv}
\rho_v = \frac{1}{2L_1\sin\theta_m}
\end{equation}

\noindent where for FLEXX, $L_0=1.75$~m is fixed by the positions of the virtual source and monochromator, whilst $L_1$ may be varied from 1.5~m to
2~m by moving the sample table along the monochromator-sample arm. It is usually kept equal to $L_0$ in order to optimise the energy resolution and beam
profile~\cite{skoulatos_rowland}.

\section{Three-Axis Mode Characterisation} \label{sec-gains}

\begin{figure}
  \begin{center}
    \includegraphics[width=\columnwidth,bb=26 180 578 607]{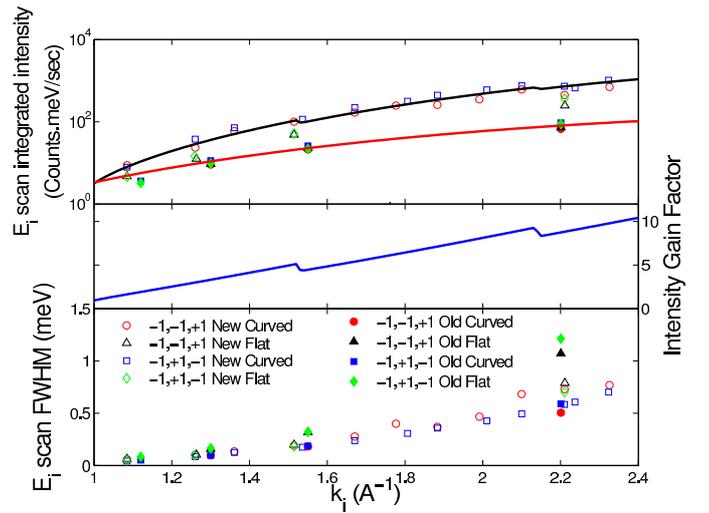}
    \caption{Fitted integrated intensities and full width at half maximum of incident energy scans of a 1~cm diameter vanadium sample. 
	     Solid lines indicate the interpolated values derived from data before (red) and after (black) the upgrade. These values were used to
	     calculate the gain factor shown in the middle panel. `Curved' indicates that the analyser was focussed and the detector was at the focal
	     distance ($L_3\approx$1~m) whilst for the `Flat' analyser, the detector was placed at its closest physical distance ($L_3\approx$0.5~m).
             $\pm1,\pm1,\pm1$ in the legend indicates the scattering senses.}
    \label{fg:vangain} \end{center}
\end{figure}

In order to characterise the intensity gains at the sample position and the energy resolution of the new FLEXX instrument, we completed a series of
measurements with a 1~cm diameter vanadium rod, scanning the incident energy, $E_i$ for a range of fixed final neutron energies, $E_f$. The data were fitted with
Gaussian peaks and the integrated intensities and full width at half maximum (FWHM) are plotted as a function of the scattered neutron wavevector $k_f$ 
in figure~\ref{fg:vangain}.

\setlength{\tabcolsep}{7pt}
\begin{table*} \renewcommand{\arraystretch}{1.0}
\begin{center}
 \begin{tabular*}{1.0\textwidth}{@{\extracolsep{\stretch{1}}}l|rr|rr|c|rr|rr|c}
  \hline 
                 & \multicolumn{5}{c|}{-1,+1,-1 Curved}                               & \multicolumn{5}{c}{-1,-1,+1 Curved}                                \\ \hline
                 & \multicolumn{2}{c|}{Old}    & \multicolumn{2}{c|}{New}    &  Gain  & \multicolumn{2}{c|}{Old}    & \multicolumn{2}{c|}{New}    &  Gain  \\ \hline
$k_i$(\AA$^{-1}$)&    Peak ct/s    &    FWHM   &    Peak ct/s    &    FWHM   &        &    Peak ct/s    &    FWHM   &    Peak ct/s    &    FWHM   &        \\ \hline 
       2.2 \ \   &       150       &   0.590   &     1056        &   0.474   &   8.8  &       126       &   0.504   &     1038        &   0.499   &   8.3  \\
       1.55      &       131       &   0.186   & \    585        &   0.155   &   5.4  &       110       &   0.182   &  \   510        &   0.170   &   4.9  \\
       1.3 \ \   &       111       &   0.097   & \    374        &   0.077   &   4.2  &       111       &   0.097   &  \   323        &   0.088   &   3.2  \\
   \hline 
  \end{tabular*}
  \caption{Vanadium incoherent elastic scans summary. FWHM denotes the energy full width at half maximum in meV at the elastic position. The gain is defined
           as the ratio $\sfrac{I_{\mathrm{new}}}{w_{\mathrm{new}}} / \sfrac{I_{\mathrm{old}}}{w_{\mathrm{old}}}$. Curved means that both the monochromator 
           and analyser were optimally focussed. In the old (pre-upgrade) measurements, the meant a vertically focussed monochromator and horizontally 
           focussed analyser. For the new measurements this means that the monochromator was focussed both horizontally and vertically, whilst the analyser 
           remained only focussed horizontally.}
  \label{tab:flux}
\end{center}
\end{table*}

\setlength{\tabcolsep}{7pt}
\begin{table*} \renewcommand{\arraystretch}{1.0}
\begin{center}
  \begin{tabular*}{0.8\textwidth}{@{\extracolsep{\stretch{1}}}l|rr|rr|rr|rr}
   \hline 
                 &  \mc{4}{c|}{-1,+1,-1 Curved}                                  &  \mc{4}{c}{-1,-1,+1 Curved}                                    \\ \hline
  Sample Size    &  \mc{2}{c|}{1 cm $\diameter$} &  \mc{2}{c|}{2 cm $\diameter$} &  \mc{2}{c|}{1 cm $\diameter$} &  \mc{2}{c}{2 cm $\diameter$}   \\ \cline{2-9}
  Virtual Source &       3cm      &     1cm      &      3cm     &      1cm       &      3cm      &      1cm      &      3cm      &     1cm        \\ \hline
$k_i$(\AA$^{-1}$)&                &              &              &                &               &               &               &                \\ \cline{1-1}
       3.0 \ \   &      1.173     &    1.073     &     1.776    &     1.540      &     1.341     &     1.269     &     1.880     &    1.601       \\
       2.2 \ \   &      0.474     &    0.442     &     0.595    &     0.559      &     0.499     &     0.480     &     0.694     &    0.646       \\
       1.55      &      0.155     &    0.151     &     0.203    &     0.189      &     0.170     &     0.159     &     0.227     &    0.212       \\
       1.3 \ \   &      0.077     &    0.072     &     0.097    &     0.086      &     0.088     &     0.083     &     0.116     &    0.110       \\
   \hline 
  \end{tabular*}
  \caption{Comparison of vanadium incoherent elastic energy widths (in meV) with different virtual source widths and sample diameters
           ($\diameter$). There is a loss of $\approx$2.5$\times$ in intensity using a 1~cm virtual source compared with the full 3~cm opening.} 
  \label{tab:vsvan}
\end{center}
\end{table*}

\setlength{\tabcolsep}{5pt}
\begin{table*} \renewcommand{\arraystretch}{0.8}
\begin{center}
  \begin{tabular*}{1.0\textwidth}{@{\extracolsep{\stretch{1}}}l|cc|cc|ccc|ccc}
   \cline{2-11} 
                 & \multicolumn{4}{c|}{Old FLEX}                                    & \multicolumn{6}{c}{New FLEXX}                       \\ \cline{2-11}
                 & \mc{2}{c|}{60-open-open}    &  \mc{2}{c|}{60-20-20}  & \mc{3}{c|}{Open Collimation}    & \mc{3}{c}{60-20-20}           \\ \hline
$k_i$(\AA$^{-1}$)&    Flat RMH     &    Flat   & Flat RMH  &    Flat    &   Curved  & Flat RMH & All Flat &  Curved & Flat RMH & All Flat \\ \hline
     2.2         &     0.0256      &   0.0245  &  0.0169   &   0.0140   &   0.0569  &  0.0344  &  0.0297  &  0.0157 &  0.0146  &  0.0140  \\
     1.55        &     0.0163      &   0.0160  &  0.0089   &   0.0084   &   0.0805  &  0.0255  &  0.0203  &  0.0095 &  0.0086  &  0.0092  \\
     1.3         &     0.0119      &   0.0127  &  0.0059   &   0.0059   &   0.0448  &  0.0191  &  0.0147  &  0.0051 &  0.0051  &  0.0049  \\
     1.12        &     0.0080      &   0.0088  &           &   0.0037   &   0.0284  &  0.0137  &  0.0094  &  0.0046 &  0.0040  &  0.0040  \\
   \hline 
  \end{tabular*}
  \caption{Momentum transfer resolution from scans of the (111) Bragg reflection of a silicon wafer. Values are the deduced full width at half 
           maximum of $|\V{Q}|$ in \AA$^{-1}$. `Curved' denotes that scans were performed with the optimum horizontal and vertical curvature 
           for the monochromator and horizontal curvature for the analyser. `Flat RMH' indicates that the monochromator
           was only focused vertically and the analyser horizontally, whilst `All Flat' shows that all curvatures were zero.} 
  \label{tab:si}
\end{center}
\end{table*}

\setlength{\tabcolsep}{7pt}
\begin{table*} \renewcommand{\arraystretch}{1.0}
\begin{center}
  \begin{tabular*}{0.9\textwidth}{@{\extracolsep{\stretch{2}}}l|c|ccc|ccc}
   \hline 
           $(hkl)$               & Energy   & \multicolumn{3}{c|}{Counts/s}   &  \multicolumn{3}{c}{FWHM}         \\
                                 & Transfer &    Old    &   3 cm    &   1 cm  &    Old    &   3 cm    &   1 cm    \\ \hline
         $(\B{0.1}~\B{0.1}~2)$   &   1.1    &    278    &    2916   &   1039  &    0.62   &   0.71    &   0.76    \\
         $(\B{0.2}~\B{0.2}~2)$   &   2.4    &\    61    &\    519   &\   183  &    0.68   &   0.74    &   0.76    \\
         $(\B{0.25}~\B{0.25}~2)$ &   3.0    &\    35    &\    302   &\   110  &    0.72   &   0.78    &   0.80    \\
         $(\B{0.3}~\B{0.3}~2)$   &   3.6    &\    20    &\    189   &\ \  66  &    0.84   &   0.91    &   0.92    \\
         $(\B{0.35}~\B{0.35}~2)$ &   4.2    &\    10    &\    112   &\ \  41  &    0.89   &   1.01    &   1.02    \\
          \hline
         $(0.1~0.1~2)$           &   1.2    &    410    &    2944   &   1267  &    0.41   &   0.54    &   0.54    \\
         $(0.15~0.15~2)$         &   1.8    &    171    &    1129   &\   462  &    0.44   &   0.57    &   0.57    \\
         $(0.2~0.2~2)$           &   2.5    &\    79    &\    573   &\   232  &    0.50   &   0.61    &   0.60    \\
         $(0.25~0.25~2)$         &   3.1    &\    45    &\    339   &\   136  &    0.57   &   0.66    &   0.65    \\
         $(0.3~0.3~2)$           &   3.6    &\    25    &\    215   &\ \  86  &    0.65   &   0.73    &   0.73    \\
         $(0.35~0.35~2)$         &   4.2    &\    14    &\    137   &\ \  53  &    0.72   &   0.85    &   0.85    \\
         $(0.4~0.4~2)$           &   4.8    &\ \   9    &\ \   81   &\ \  31  &    0.58   &   0.85    &   0.86    \\
  \hline 
  \end{tabular*}
  \caption{Summary of inelastic count rate gains and resolution broadening deduced from measurements of a transverse acoustic phonon in a single
           crystal of lead at room temperature. Energy transfer and full width at half maximum (FWHM) of the inelastic peaks are in meV. The 
           final neutron wavevector was fixed at $k_f=$1.55~\AA$^{-1}$, a single 60' Soller collimator was used after the monochromator 
           in the pre-upgrade measurements, whilst measurements after the upgrade used no collimation. All measurements employed the `chair' 
           configuration with scattering senses, $-1,-1,+1$. In the case of the upgraded instrument, an open virtual source (3cm) as well as
           slightly closed source (1cm) were used.}
  \label{tab:pb}
\end{center}
\end{table*}

Since the dataset for the upgraded instrument is much more extensive than for the old FLEX, the intensity gain, $I_{\mathrm{new}}/I_{\mathrm{old}}$,
was calculated by interpolating between the data points, and is generally linear in $k_f$, rising from $\approx$1 at $k_f\approx$1~\AA$^{-1}$ to
$\approx$10 at $k_f\approx$2.4~\AA$^{-1}$. Steps in the gain factor arise from scattering by the larger thicknesses of aluminium comprising the
windows between different guide sections in the upgraded instrument compared to the old guide system.

The bottom panel of figure~\ref{fg:vangain} shows the measured energy FWHM for elastic incoherent scattering from Vanadium. It shows that the
resolution of the new FLEXX is slightly improved compared with the old instrument. Representative measurements of the peak counts per second, energy
full width at half maximum and gain factors for some incident wavevectors are summarised in table~\ref{tab:flux}, where we have defined the gain
factor as the ratio $\sfrac{I_{\mathrm{new}}}{w_{\mathrm{new}}} / \sfrac{I_{\mathrm{old}}}{w_{\mathrm{old}}}$ of the count rate of the new instrument
over the old multiplied by the ratio of the energy width of the old over the new.

\begin{figure}
  \begin{center}
    \begin{tabular}{@{\extracolsep{\fill}}cc}
      \includegraphics[width=0.43\columnwidth,bb=43 183 551 587]{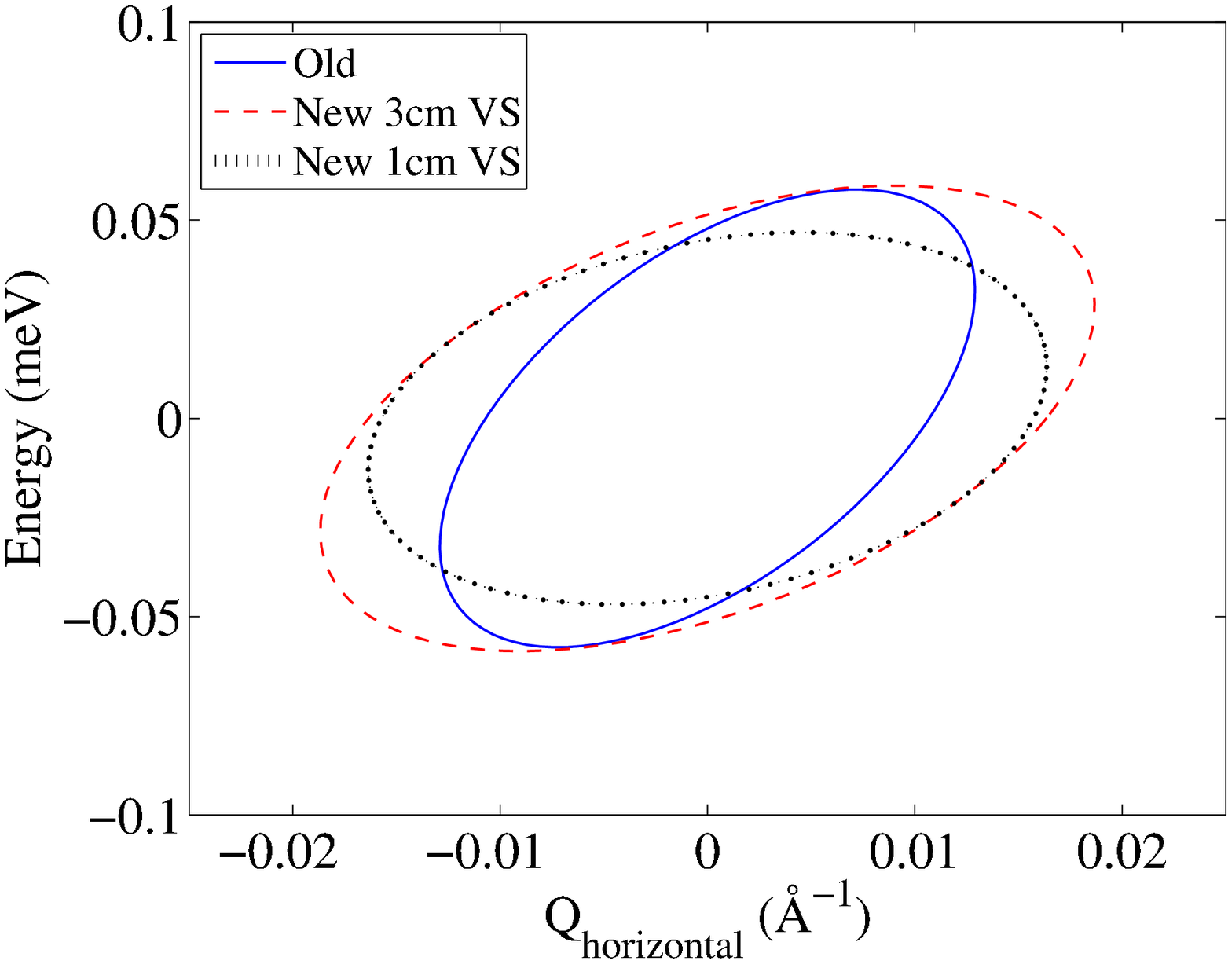} &
      \includegraphics[width=0.48\columnwidth,bb=22 183 561 608]{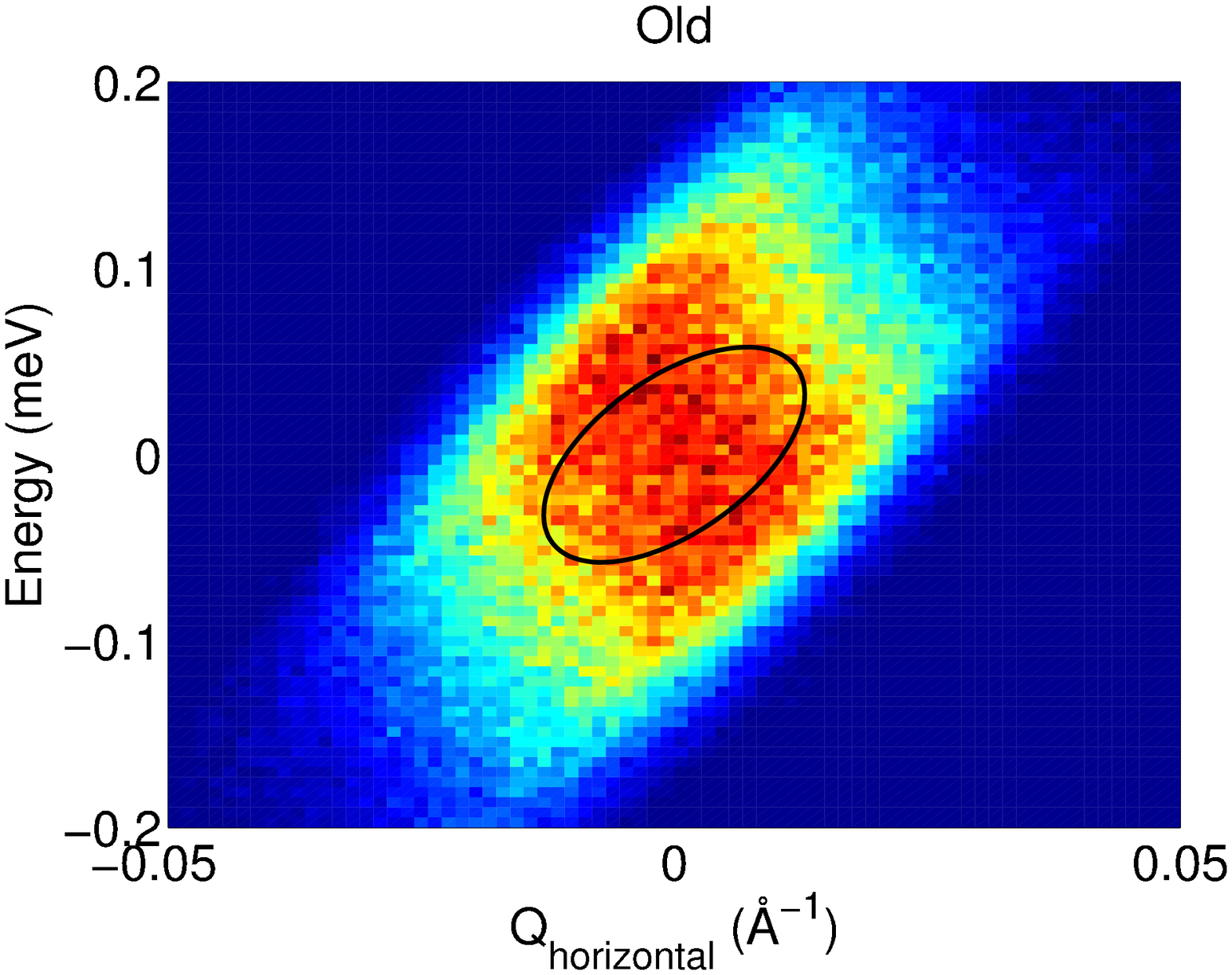} \\
      \includegraphics[width=0.48\columnwidth,bb=22 183 561 608]{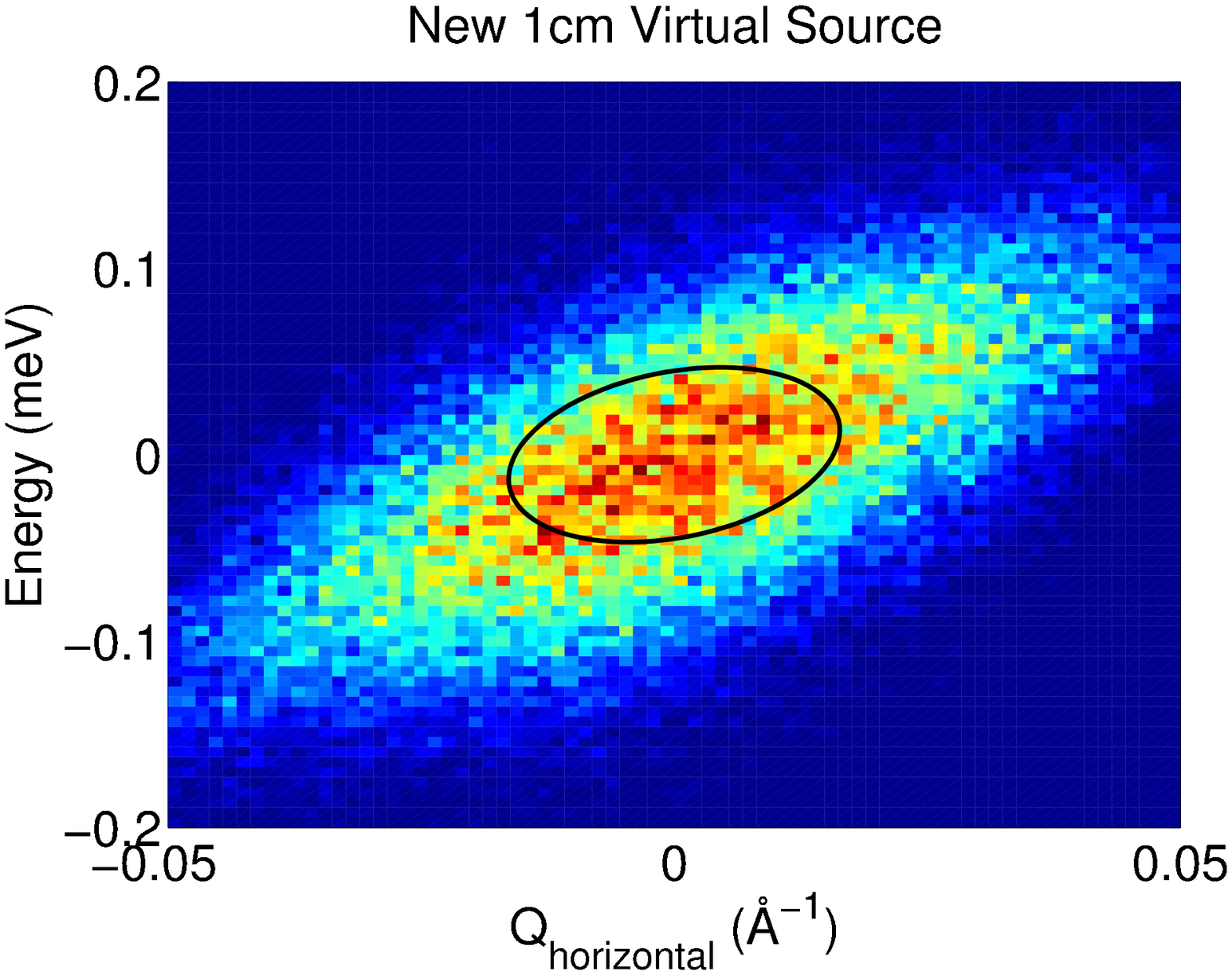} &
      \includegraphics[width=0.48\columnwidth,bb=22 183 561 608]{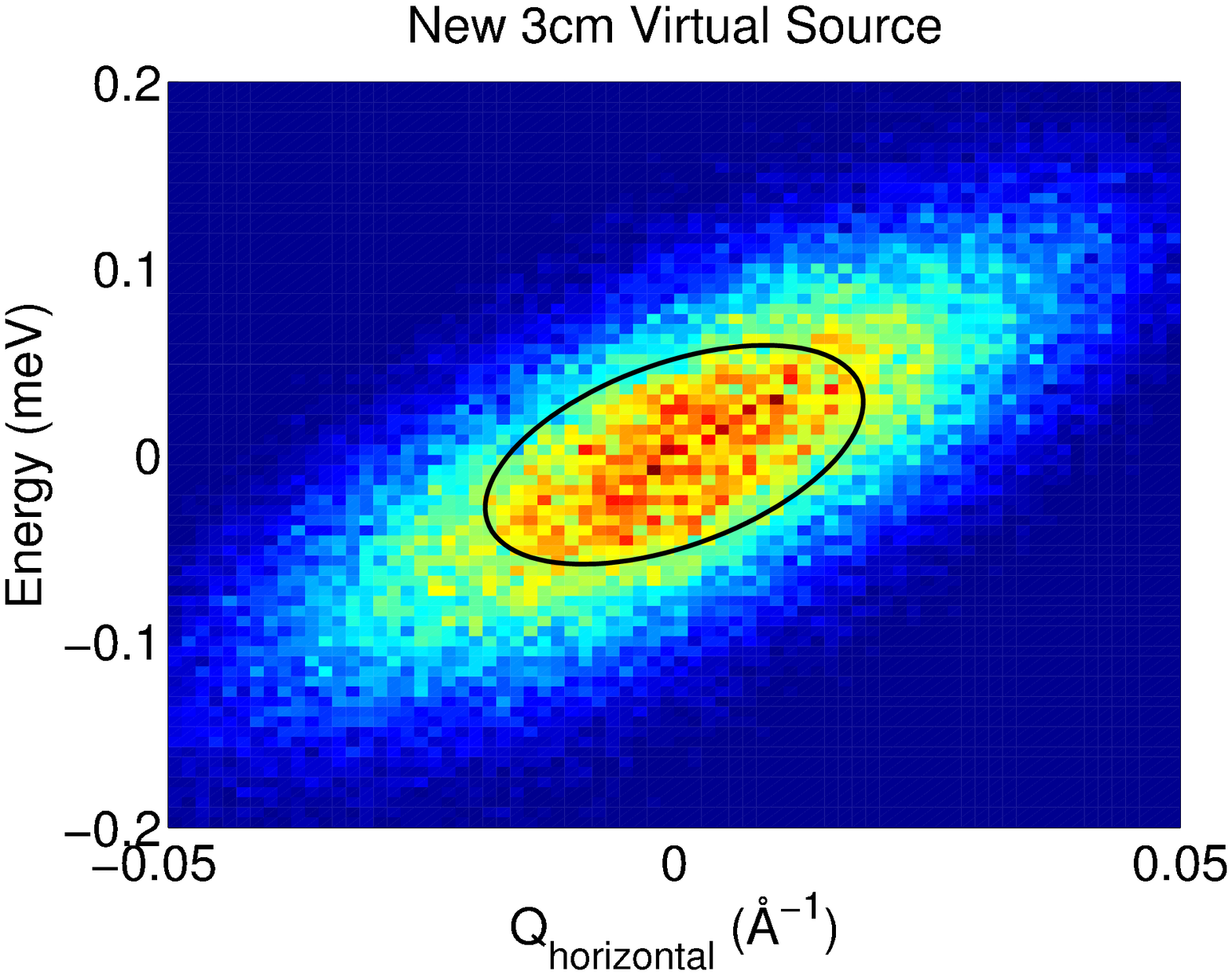}
    \end{tabular}
    \caption{Resolution volumes calculated using Popovici's method (top left and solid ellipses in other panels) and Monte Carlo ray tracing in McStas
             (color plots). $k_i$=$k_f$=1.55~\AA$^{-1}$ with scattering sense (-1,+1,-1) for all cases. Simulations for the old FLEX (top right panel)
	     used a 60' collimator after the monochromator with all other collimation open. For the new FLEXX (bottom panels), no collimation was
	     used. All applicable monochromator and analyser focussing was used (vertical monochromator, horizontal analyser for old FLEX; double
             focussing monochromator, horizontal analyser for new FLEXX).}
    \label{fg:resellip} \end{center}
\end{figure}

The slightly improved energy resolution for the new guide system may be explained by comparing the resolution volumes calculated in
figure~\ref{fg:resellip} using both Monte Carlo ray-tracing and with the analytical method of Popovici~\cite{popovici75}. The top and bottom colour
plots on the right side of the figure compares the resolution volume for the old and new guide system and shows clearly that the energy width of the
new instrument is narrower than for the old. This is not reflected in the analytical calculation, shown in the top left panel, however. It may be that
the coupling between the divergence of the neutron and their wavelength spread introduced by horizontally focussing the
monochromator~\cite{cussen_phaseE} is not fully modelled by Popovici's method, which folds the effect of the curvature into that of the crystal
mosaic.

In principle, horizontally focussing neutrons from a point source onto a point sample with a {\it zero mosaic} monochromator according to
equation~\ref{eq:rmh} would result in a perfectly monochromatic beam at the sample, if the distances between source and monochromator ($L_0$) and
monochromator and sample ($L_1$) are equal~\cite{cussen_phaseE}. An extended neutron source can be thought of as a series of point sources along a
line, each of which would be imperfectly imaged by the focussed monochromator onto different positions with slightly different energies, resulting
both in a broader spatial extent of the beam at the sample position and greater wavelength spread. Whilst in practice, the mosaicity must be non-zero
and fixed and thus imposes a minimum beam width and wavelength spread, FLEXX was designed with a horizontal diaphragm, placed at the focus of the
elliptical guide and at $L_0$=1.75~m from the center of the monochromator, acting as a virtual neutron source. By reducing the size of this virtual
source, better focussing may be achieved, resulting in a more monochromatic and narrower beam.
Monte Carlo simulations~\cite{habicht_skoulatos_virtualsource} showed that by changing the virtual source width from 3~cm (the guide width at the
focus of the elliptical guide) to 1~cm, the energy resolution may be improved by up to 40\%. Finally, the sample size also affects the measured
energy resolution, since it is focussed by the analyser onto the detector, and thus acts like a source.

In order to check the effect of the virtual source size and sample width on the measured energy resolution, we performed a series of measurements
at different incident energies of the incoherent elastic width from two vanadium samples: a solid rod, 1~cm in diameter, and a hollow cylinder 2~cm in
outer diameter. The results are summarised in table~\ref{tab:vsvan}. The measurements do show that smaller samples and smaller virtual source sizes
results in narrower energy widths. However, the improvement in energy resolution from reducing the virtual source from 3~cm to 1~cm is only of the
order of 10\%, at the cost of a $\approx$2.5$\times$ reduction in count rate. The effect is more pronounced for larger samples and higher incident
energies.

Finally, we note that the measurements in the upgraded instrument were made with the monochromator horizontally focussed and without any Soller
collimators. This is in contrast to measurements before the upgrade where a Soller collimator, $\alpha_1=$60', after the (horizontally flat)
monochromator was often used, in order to obtain a cleaner spatial beam profile. Whilst this results in a gain in count rate, it is balanced by a
corresponding broadening of the momentum resolution as can be seen from table~\ref{tab:si}. This in turn results in larger measured energy widths for
dispersive excitations, such as phonons, as shown in table~\ref{tab:pb}. For dispersionless excitations, we anticipate that the energy resolution
should not change, or may indeed have improved as shown by measurements of energy widths of the incoherent elastic line in table~\ref{tab:flux}.

The momentum $\V{Q}$ resolution of the instrument was determined by $[hhh]$ scans over the (111) Bragg reflection of a Silicon wafer, under the
assumption that the mosaic of the silicon wafer is less than the instrument resolution. Furthermore, with both the monochromator and analyser
horizontally focused and no collimation, the scans showed a multiple peak (up to 5 peaks could be discerned in many cases) structure, as the different
blades of the analyser (which consists of 15 1$\times$10~cm PG(002) blades with 0.6$^{\circ}$ mosaic that may be horizontally curved) reflect the 
scattered beam in turn.\footnote{For $k_F$=1.55\AA$^{-1}$, the analyser scattering angle is $\theta_A=37.2^{\circ}$, and the optimum curvature is 50.3~cm$^{-1}$. 
This implies that the blades either side of the central blade are rotated by $\Delta\theta_A=0.01/(1/0.503)=\pm 0.3^{\circ}$ which is approximately the
same as the angle between these blades and the central blade as seen by the sample which is 1.4~m from the analyser $\Delta\theta_S=0.01/1.4=\pm 0.4^{\circ}$,
thus satisfying the Bragg condition if the sample is rotated by that slight amount.}
The effect is greatly reduced (to a double peak structure) if either monochromator or
analyser are left horizontally flat. A single, sharp peak results if both analyser and monochromator are flat, although the measured $\V{Q}$
resolution in this case is still worst than for the pre-upgrade instrument by approximately 20\%.  In contrast with tight collimation, the measured
$\V{Q}$ resolution is the same as previously, whilst the count rate is approximately 40\% higher, which is due to improvements in the cold source.
Finally we note that the mosaic of the silicon wafer is very small, and for usual single crystal samples with mosaic of the order of 1$^{\circ}$ the
multiple peak structure does not manifest when both monochromator and analyser are horizontally focussed.

\begin{figure}
  \begin{center}
    \includegraphics[width=\columnwidth,bb=31 179 545 587]{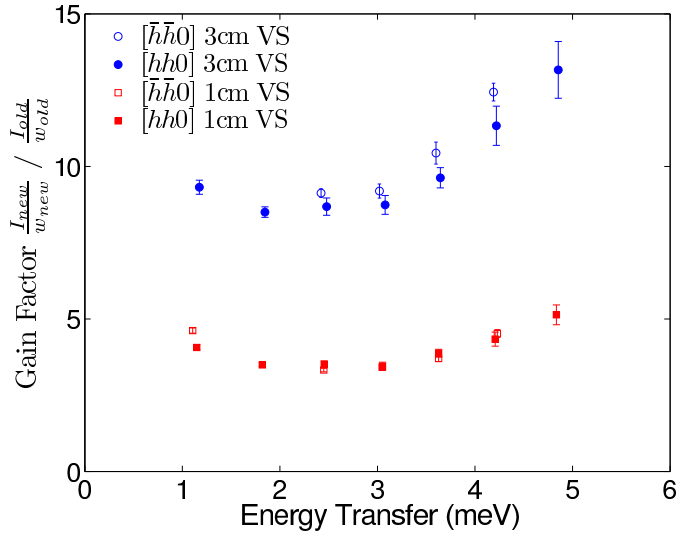}
    \caption{Gain factor derived from fitted integrated intensities and FWHM of lead phonons as a function of energy transfer in the anti-W 
             (long-chair) configuration, at fixed final wavevector of 1.55~\AA$^{-1}$. No collimation were
             used in the new measurements on the upgraded FLEX, whereas the old data was obtained with 60' collimation after the
             monochromator (open-60'-open-open).}
    \label{fg:phongain} \end{center}
\end{figure}

The inelastic intensity gains and energy resolution were determined by measuring the dispersion of a transverse acoustic phonon branch in the [220] direction
around the Bragg peak at (002) in a large single crystal of lead. The measurements were fitted to a Gaussian peak shape and the ratio of the
integrated intensities, FWHM and background between the old and upgraded instruments at different wavevector transfer $q$, and hence energy transfer,
are plotted in figure~\ref{fg:phongain}, and summarised in table~\ref{tab:pb}. We restrict ourselves to low energy transfers such that a linear
dispersion obtains. Open symbols in the figure indicate that $q$ is parallel to $[\bar{1}\bar{1}0]$, whereas filled symbols are from measurements
with $q$ along [110]. 

The data shows an approximate gain factor of 10$\times$ in intensity at low energy transfer, rising at higher energies, due to the $m=3$ supermirror
guide which better transports higher energy neutrons than the previous $^{58}$Ni guide. Using a smaller virtual source had little effect on the
measured energy resolution in this case, as shown by the FWHM in table~\ref{tab:pb}. This is because the much coarser momentum resolution of the
upgraded instrument means that a larger part of the dispersion falls within the resolution ellipsoid. The measured energy widths in a constant-$\V{Q}$
scan is then much more a function of the momentum resolution and the dispersion. This is especially true here for a steeply dispersing acoustic
phonon.

This can be seen by comparing the bottom panels of figure~\ref{fg:resellip} which shows a resolution volume for the upgraded instrument for a 1~cm and
a 3~cm virtual source. This demonstrates that whilst reducing the virtual source results in a somewhat narrower energy width, the reduction in the
momentum width is less pronounced. Furthermore, the momentum width of the upgraded instrument in both cases is much larger than that of the old, so
that one should expect the measured phonon energy widths to be larger also. A close examination also shows that relatively more neutrons are accepted
with larger momentum divergence in the case of the 1~cm source than the 3~cm, with the acceptance volume having a more parallelepiped than ellipsoidal
shape. This may account for the curious observation of the slightly larger energy FWHM in this case compared to the 3~cm source, as shown in table~\ref{tab:pb}.

\section{Conclusions} \label{sec-conc}

We have described the performance of the upgraded cold neutron triple-axis spectrometer FLEXX at the Berlin Neutron Scattering Center. We find gains
of over 10$\times$ in the count rate for inelastic excitations at the cost of coarser momentum resolution resulting from the improved guide system and
a larger double focussing monochromator. For non-dispersive modes, we also found a slight improvement in the energy resolution. The use of a velocity
selector has removed the need for a neutron filter and allows much greater flexibility in choosing the incident and final neutron energies, whilst the
$m=3$ supermirror guides allows measurements into the thermal energy range, up to $\approx$20~meV incident energy without much reduction in neutron
flux. Furthermore, the measured background has not increased noticeably from the old instrument, at approximately 2~counts/minute. The non-neutron
background, measured with a 5~mm thick borated plastic piece blocking the beam between the sample table and analyser, is $\approx 1$~count per 10
minutes.

We thus expect the improved FLEXX spectrometer to be an important tool in the study of low energy excitations in condensed matter physics.
Further upgrades to the polarised neutron, resonance spin echo and flatcone secondary spectrometer options will be described in forthcoming
publications.


 \bibliographystyle{elsarticle-num} 

\end{document}